\documentclass[aps,pra,twocolumn,showpacs,groupedaddress]{revtex4}  
\usepackage{amsmath}
\usepackage{tipa}
\usepackage{bbm}
\usepackage{txfonts}
\usepackage{graphicx}
\usepackage{dcolumn}
\usepackage{bm}
\usepackage{amssymb}
\usepackage{latexsym}
\usepackage{color}
\begin{document}


\title{Generalized Bloch oscillations of ultracold lattice atoms subject to higher-order gradients}
\author{Qian-Ru Zhu}
\author{Shou-Long Chen}
\author{Shao-Jun Li}
\author{Xue-Ting Fang}
\author{Lushuai Cao}\email[E-mail: ]{lushuai_cao@hust.edu.cn}
\author{Zhong-Kun Hu}\email[E-mail: ]{zkhu@hust.edu.cn}

\affiliation{MOE Key Laboratory of Fundamental Physical Quantities Measurement $\& $
Hubei Key Laboratory of Gravitation and Quantum Physics, PGMF and School of Physics,
Huazhong University of Science and Technology, Wuhan 430074, P. R. China}

\date{\today}

\begin{abstract}
  The standard Bloch oscillation normally refers to the oscillatory
  tunneling dynamics of quantum particles in a periodic lattice plus a
  linear gradient. In this work we theoretically investigate the
  generalized form of the Bloch oscillation in the presence of additional
  higher order gradients, and demonstrate that the higher order gradients
  can significantly modify the tunneling dynamics, particularly in the
  spectrum of the density oscillation. The spectrum of the standard Bloch
  oscillation is composed of a single prime frequency and its higher
  harmonics, while the higher-order gradients in the external potential
  give rise to
  fine structures in the spectrum around each of these Bloch frequencies,
  which are composed of serieses of frequency peaks.
  Our investigation leads to a two-fold consequence to the applications
  of Bloch oscillations for measuring external forces:
  For one thing, under a limited resolution of the measured spectrum,
  the fine structures would manifest as a blur to the spectrum,
  and leads to intrinsic errors to the measurement.
  For another, given that the fine structures could be experimentally
  resolved, they can supply more information of the external force
  than the strength of the linear gradient, and be used to measure more complicated forces.
\end{abstract}


\maketitle

\setlength{\arraycolsep}{0.8pt}

\section{INTRODUCTION} \label{section:I}
Ultracold atoms, owing to their perfect isolation to the environment and flexible tunability,
have become an ideal platform for fundamental physics research and practical
applications \cite{Blo08,Jak05}. Among these investigations, a paradigm is the Bloch oscillation
(BO) of ultracold lattice atoms, which not only reveals the wave natures of quantum particles,
but also finds various applications, $e.g.$ in precision measurements. BO, known as the periodic
oscillation of a particle subject to a periodic potential plus a liner gradient,
was firstly introduced in solid-state systems \cite{Zen34}, and have been experimentally
explored in various ultracold atomic ensembles, such as with Bose-Einstein
condensates \cite{Dah96,Mor01,Pri98,Bat04,Cla06,Fer06,Gus08,Gei18,Fat08},
degenerate Fermi gases \cite{Roa04}, and strongly correlated lattice atoms \cite{Pre15}.
BO of ultracold atoms has been theoretically proposed
to be applied for the measurement of the gravity acceleration \cite{Cla05},
the nonlinear interaction between atoms \cite{Bre07}, as well as forces at short
distances \cite{Car05,Sor09}, such as the Casimir force and even the non-Newtonian
gravity forces. In experiments, the high precision measurements of
$h/m_{Rb} $ \cite{Bat04,Cla06}, and the gravity acceleration \cite{Fer06,Gus08,Gei18,Fat08,Roa04}
have also been performed.

The standard BO can also be generalized to more complicated setups, which bring in new effects
and applications. For instance, temporal modulations of the lattice
potential \cite{Wan04,Tar12,Iva08,Hal10,Pol11} or the interaction strength \cite{Dia13}
have been demonstrated to induce the so-called super Bloch oscillations, which can be used to
transport atoms in the lattice with a controllable manner. The BOs in nonuniform lattices,
such as aperiodic lattices \cite{Wal10,Dug16}, disorder lattices \cite{Sch08} as well as zigzag
and helix lattices \cite{Sto15} have also been extensively investigated. In this work,
we consider an alternative generalization scheme of BO in optical lattices, in which higher
order gradients are taken into account besides the linear one. This generalization,
termed as generalized Bloch oscillation (GBO), is in favor of the realistic setups of optical
lattices in the presence of external forces, such as gravity or Casimir force, which contains
not only the linear but also higher order gradients. We demonstrate that, on the one hand,
in the presence of the higher order
gradients, the dynamics of the lattice atoms still maintains an oscillatory
behavior. On the other hand and more importantly, the frequency spectrum of GBO is significantly
modified and the higher order gradients give rise
to the fine structures in the spectrum, manifested as
splitting of the Bloch frequencies, $i.e.$ the prime frequency and its
higher harmonics, into series of frequency peaks.
The fine structures present a strong influence on
the measurement of external forces based on the tunneling dynamics
of lattice atoms.
For one thing, in the measurements of the linear forces by the Bloch
frequencies, the fine structure induced by the residual higher order
gradients, broadens the measured frequency peaks of the corresponding
Bloch frequencies under a limited experimental resolution of the spectrum
and leads to intrinsic errors to the measurements.
For another, provided an improved resolution
of the spectrum to identify the fine structure, it is possible to decode
the strength of the higher-order gradients from the fine structure, and
holds the potential in precision measurement of complicated forces.

This paper is organized as follows. In Sec. \ref{section:II} we present the Generalized Bloch
oscillation in optical lattices with quadratic gradient, including the setup (\ref{secIIA}),
analytical results (\ref{secIIB}) and
numerical verifications (\ref{secIIC}). A brief discussion are given in Sec. \ref{section:III}.

\section{GENERALIZED BLOCH OSCILLATION IN OPTICAL LATTICES WITH QUADRATIC GRADIENT} \label{section:II}
\subsection{Setup}\label{secIIA}
In this work, we consider the quantum dynamics of ultracold atoms confined in one-dimensional
optical lattices in the presence of an external potential, containing gradients of different
orders.
The interaction between atoms is set to be zero, which can be experimentally realized either by
tuning the interaction to approaching zero via Feshbach resonance \cite{Gus08,Gei18,Fat08},
or directly loading single atoms to the optical lattices \cite{Pre15}. Under the condition of
non-interacting lattice atoms, the setup is reduced to a single-particle one, with the Hamiltonian
given as:
\begin{equation}
H=-\frac{{{\hbar }^{2}}}{2M}\partial _{x}^{2}+{{V}_{l}}\left( x \right)+{{V}_{ex}}\left( x \right)\ ,\label{Eq:1}
\end{equation}
where $M $ denotes the mass of the atoms,
${{V}_{l}}\left( x \right)={{V}_{0}}{{\sin }^{2}}\left( kx \right)\ $ is the lattice potential,
and ${{V}_{ex}}\left( x \right)\ $ refers to the external potential. The external potential can
be decomposed as a summation of gradients of different orders, $i.e.$
${{V}_{ex}}=\sum_{\alpha\geq 1}{{{V}_{\alpha }}{{x}^{\alpha }}}\ $,
where, $e.g.$, $\alpha= $1 and 2 refers to the linear and quadratic gradients, respectively.

In our investigations, the atoms are initially loaded to a single site, and the dynamics is mainly
around this initial site. In the following, we shift the origin of the coordinate to the potential
minimum of the initial site, denoted as ${{x}_{0}}\ $, and the external potential becomes
$V_{ex}^{0}\left( x \right)=\sum\limits_{\alpha \ge 1}{{{V}_{\alpha }}{{\left( x+{{x}_{0}} \right)}^{\alpha }}}\equiv \sum\limits_{\alpha \ge 0}{{{{\tilde{V}}}_{\alpha }}{{x}^{\alpha }}}\ $
in the new coordinate, in which
${{\tilde{V}}_{\alpha }}=\sum\limits_{\beta \ge \alpha }{C_{\beta }^{\alpha }}{{V}_{\beta }}x_{0}^{\beta -\alpha }\ $, with $C_{\beta }^{\alpha }\ $
denoting the binomial coefficient. Alternatively,
$V_{ex}^{0}\left( x \right)$ is equivalent to the Taylor expansion of
the given external force around the initial site, and
${{\tilde{V}}_{\alpha }}$ coincides to the corresponding expansion coefficient.
By carefully choosing $\tilde{V}_\alpha$,
$V_{ex}^{0}\left( x \right)$
can effectively model a wide range of external forces, such as the Casimir force and the van der
Waals force, and our investigations
based on $V_{ex}^{0}\left( x \right)$ can be directly
applied to these cases.

Before proceeding to the GBO, let us recall the main characteristics of the standard BO.
In the standard BO that the atoms are initially loaded to a single site of an optical lattice
in the presence of a linear gradient, $i.e.$
${{V}_{\alpha }}={{V}_{1}}\delta \left( \alpha ,1 \right)\ $,
the atoms undergo a breathing-type periodic oscillation around the initial site,
and the spectrum of BO is composed of a single prime frequency and its higher harmonics.
The prime frequency ${{\omega }_{B}}=\pi {{V}_{1}}/\hbar k \ $ linearly depends on the strength
of the gradient, which stimulates various applications in precision measurements
\cite{Bat04,Cla06,Fer06,Roa04,Cla05,Bre07,Car05,Sor09}. In this work we extend to investigate
the generalized Bloch oscillation (GBO) in the presence of higher order gradients of the external
 potential, with a focus on the spectrum of GBO.
\subsection{Analytical results}\label{secIIB}
\subsubsection{GBO under a quadratic tilt}
We firstly consider the simplest case of GBO, with the external potential composed of a linear and
a quadratic gradient,
$i.e.$ $V_{ex}^{0}\left( x \right)=
{{\tilde{V}}_{0}}+{{\tilde{V}}_{1}}x+{{\tilde{V}}_{2}}{{x}^{2}}\ $,
which we term as the quadratic GBO. The dynamics of the lattice atoms is demonstrated by the
one-body density oscillation of each site
${{\rho }_{i}}\left( t \right)\text{=}\int\limits_{x\in i-th\text{ site }}{dx\left\langle
\Psi \left( t \right) \right|\hat{x}\left| \Psi \left( t \right) \right\rangle }\ $,
as well as the corresponding frequency spectrum, ${{\rho }_{i}}\left( \omega  \right)\ $,
where ${i} $ indexes the site of lattice. Our main results can be summarized as follows:
For one thing, the one-body density ${{\rho }_{i}}\left( t \right)\ $ of quadratic GBO still
maintains a periodic structure. For another, and more importantly, a major difference between
the standard BO and the quadratic GBO arises in the spectrum. In the case of the standard BO,
the spectrum ${{\rho }_{i}}\left( \omega  \right)\ $ is composed of peaks located at
${{\omega }_{B}}\ $ and its higher harmonics, while the additional quadratic gradient splits
each of these Bloch frequencies into a series of equidistant peaks. Taking the equidistant
peaks around ${{\omega }_{B}}\ $ for instance, the location of these peaks can be derived as
\begin{equation}
\omega \left( n \right)=\,\ \frac{{{{\tilde{V}}}_{1}}}{\hbar }\frac{\pi }{k}+\left( 2n-1\right)\frac{{{{\tilde{V}}}_{2}}}{\hbar }{{\left( \frac{\pi }{k} \right)}^{2}}\ .\label{Eq:2}
\end{equation}
In equation (\ref{Eq:2}), $n$ labels the peaks around the dominant frequency
${{\omega }_{B}} $ and runs over all the integers.
Equation (\ref{Eq:2}) shows that these equidistant peaks are centered at
${{\omega }_{B}}={\pi {{{\tilde{V}}}_{1}}}/{\hbar k}\ $, and the spacing between the neighbor
peaks is
\begin{equation}
\delta \omega ={2{{\pi }^{2}}{\tilde{V}_{2}}}/{\hbar {{k}^{2}}}\ .\label{Eq:3}
\end{equation}
Shown from the above equations, the centered value and the spacing of the equidistant peaks are
linearly dependent on the strength of the linear and the quadratic gradients, respectively.
The derivation of the above equations is done by applying the first-order perturbation theory,
in which the quadratic term is taken as the perturbation to the Wannier-Stark
states \cite{Glu02,Wan60}. The detailed derivation can be found in the appendix.

Equations (\ref{Eq:2}) and (\ref{Eq:3}) indicate that in the use of BO for measuring
the linear gradient, the residual of the quadratic term can splits the prime frequency into
a series of peaks. In the case that the fine structure composed of the split
peaks could not be well resolved in the measured spectrum, it
behaves as the broadening of the measured profile of the prime frequency,
and consequently induces errors to the measurement.
On the other hand,
given that the fine structure of the equidistant peaks could be resolved
in experiments,
we suggest that the quadratic GBO could be used to realize the simultaneous measurements of
the linear and quadratic gradients of an external potential.
The measurement scheme involves performing GBO of lattice atoms, and
extracting the strength of the linear and the quadratic gradients
from the averaged value and the spacing of the equidistant peaks, respectively. Comparing
to the differential measurement strategy of the
quadratic gradient, which requires measuring the external
force at different locations, the GBO can lead to an $\textit{in-situ} $ and simultaneous
measurement of the linear and quadratic gradients,
which could be particularly useful in situations within short distances.
\subsubsection{GBO under a general tilt} \label{section:B2}
\begin{figure*}[htp]
\includegraphics[width=1.0\textwidth]{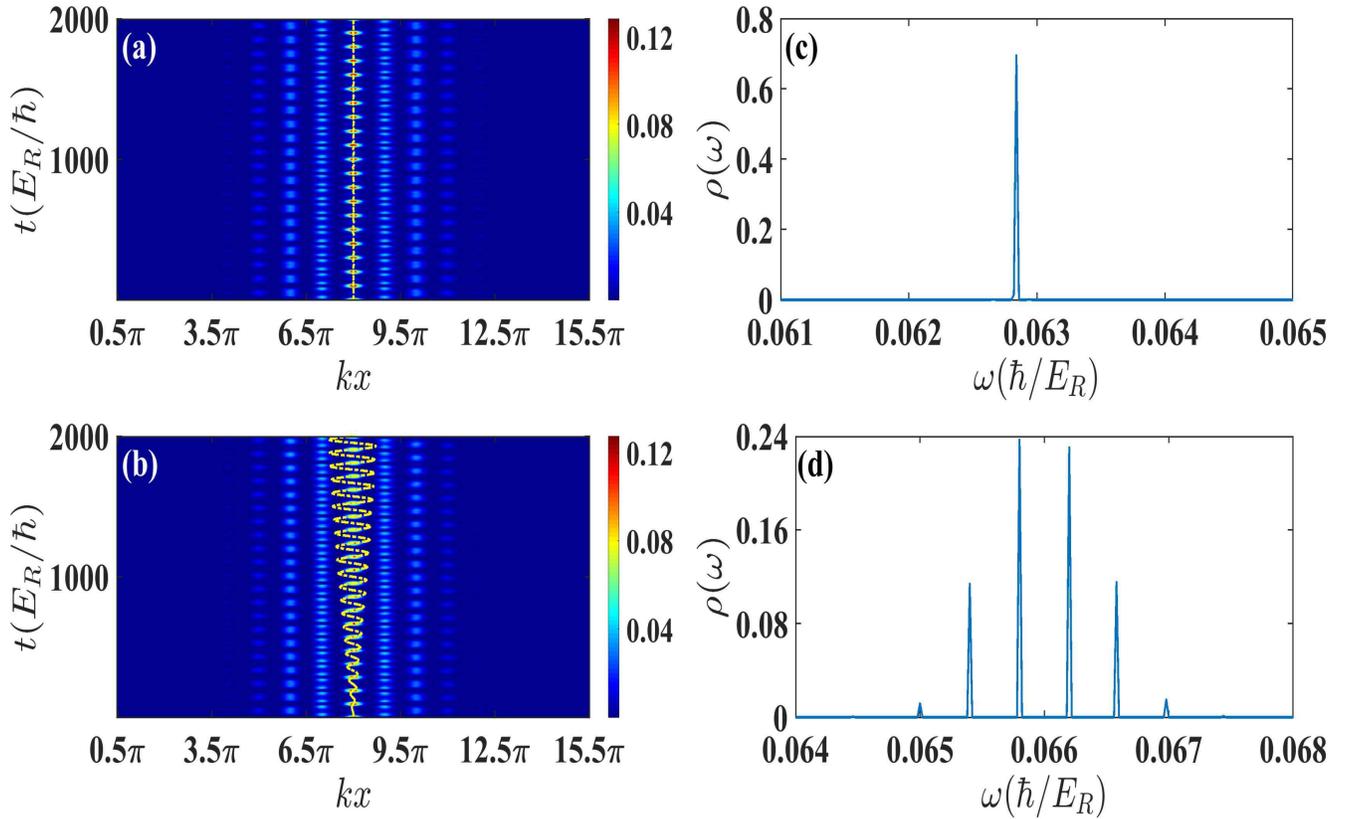}
\caption{\label{fig:2}
Comparison of the standard BO and the quadratic BO in terms of the density
distribution ((a) and (b)), as well as the density spectrum ((c) and (d)).
In figures (a) and (b), the center-of-mass is shown with the yellow dashed lines.
In the figures, ${{E}_{R}}\equiv \frac{{{\hbar }^{2}}{{k}^{2}}}{2M}\ $ refers to the recoil energy.
Taking the (G)BO of Potassium atoms in a lattice of
wavelength $\lambda=873nm$ for example, the time unit $E_R/\hbar$ is roughly $2.4\times10^{-5}s$
and the corresponding frequency unit
$\hbar/E_R\simeq 4.1\times 10^{4}Hz$.}
\end{figure*}

We turn to consider the general case of the external potential, which
is composed of arbitrarily many gradients of different orders, that¡¯s
$V_{ex}^{0}\left( x \right)=\sum_{\alpha\geq 0 }{{{{\tilde{V}}}_{\alpha }}{{x}^{\alpha }}}\ $.
Under the condition that the linear gradient dominates over the higher order terms in the external
potential, the spectrum ${{\rho }_{i}}\left( \omega  \right)\ $ of the corresponding GBO also
presents multiple series of frequency peaks, located around each
of the Bloch frequencies $n\times{{\omega }_{B}}\ $. The equidistant behavior in the quadratic GBO,
however, breaks, and the frequency of the peaks around ${{\omega }_{B}}\ $ becomes:
\begin{equation}
\omega \left( n \right)=\frac{1}{\hbar }\sum\limits_{\alpha \ge 1}{{{D}_{\alpha }}\left( n \right){\tilde{V}_{\alpha }}}\ ,\label{Eq:4}
\end{equation}

\begin{equation}
{{D}_{\alpha }}={{\left( \frac{\pi }{k} \right)}^{\alpha }}\sum\limits_{l=0}^{\alpha }{\left\{ C_{\alpha }^{l}\left[ {{n}^{\alpha -l}}-{{(n-1)}^{\alpha -l}} \right]\left\langle {{\delta }^{l}} \right\rangle  \right\}} .\label{Eq:5}
\end{equation}
In the above equations
$\left\langle {{\delta }^{l}} \right\rangle
=\sum_{\delta }{J_{\delta }^{2}}
\left( \frac{2tk}{{{{\tilde{V}}}_{1}}\pi } \right){{\delta }^{l}}\ $,
where ${{J}_{\Delta }}\left( x \right)\ $ is the $\Delta \ $-th order Bessel function of the first kind,
$t\ $ refers to the hopping strength between neighbor sites of the
lattice with no external potential, and the summation
runs over all integers.
\begin{figure}[htb]
\includegraphics[width=0.5\textwidth]{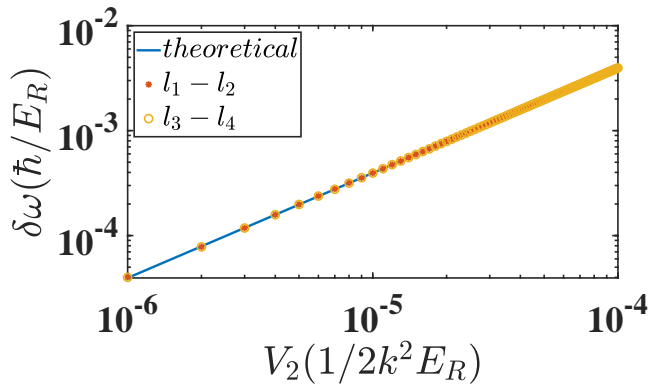}
\caption{\label{fig:3} The comparison of the neighbor peaks' spacings
, with the analytically derived value. Red star and orange circle denote
the frequency spacing between the first and last two peaks, respectively. Blue solid line denotes the theoretical value.}
\end{figure}

\begin{figure*}[htb]
\includegraphics[width=1.0\textwidth]{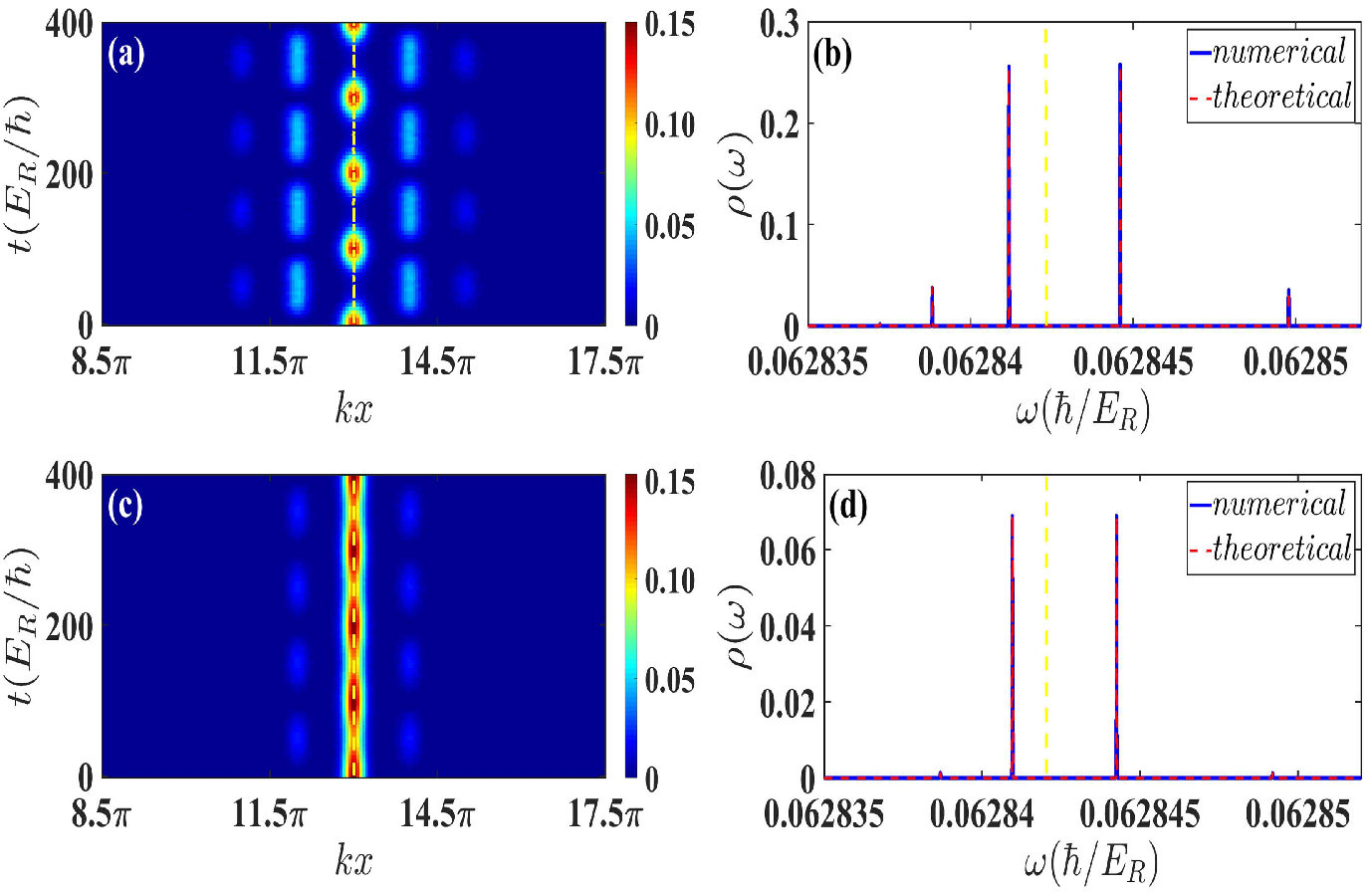}
\caption{\label{fig:4} Figures (a) and (b) present the time evolution of the density distribution ${{\rho }_{j}}\left( t \right)\ $
and the corresponding spectrum of the GBO in the presence of
the thermal Casimir-Polder plus the gravity potential,
where lattice heights are $8 E_R$.
Figures (c) and (d) plot ${{\rho }_{j}}\left( t \right)\ $
and the spectrum for lattice height $12 E_R$.
In figures (a) and (c), the center-of-mass is shown with the yellow dashed
lines. In figures (b) and (d),
the prime frequency of the standard BO, taking
into account the gravity force and the linear component of the thermal
force is marked by the yellow dashed lines, in order to compare to the
exact fine structure of the related GBO. }
\end{figure*}

The results given in equations (\ref{Eq:4}) and (\ref{Eq:5}) again
demonstrates that
the higher-order gradients generate fine structures in the spectrum, which
encodes the detailed information of the composed external potential.
In the case that the linear gradient is the dominant term of the external
potential and the higher-order gradients behaves as a perturbation,
the fine structure can induce errors in measuring the linear gradient
with the standard BO.
Moreover, the results also indicate that the characteristic frequencies of GBO still encodes the
information of the external forces, in terms of the
strength of the gradients of different orders, which could be explored
for measuring forces containing multiple gradients. It should also be
pointed out that in equations (\ref{Eq:4}) and (\ref{Eq:5}), the locations of the
frequency peaks become dependent on the lattice parameters, such
as the hopping strength, which
requires careful choice and calibrations of these parameters for applications based on GBO.
\subsection{Numerical verifications}\label{secIIC}
In this section we present numerical results on the GBO, in terms of the temporal one-body density
oscillation $\rho \left( x,t \right)\ $ and the corresponding integrated spectrum
$\rho \left( \omega  \right)\text{=}\sum\limits_{i}{{{\rho }_{i}}\left( \omega  \right)}\ $,
where ${{\rho }_{i}}\left( \omega  \right)\ $ denote the density spectrum of the ${i} $-th site, and the summation of $i$ runs over all the
occupied sites during the tunneling process.
Since we focus on the non-interacting system, our simulation reduces to a single-particle problem,
in which the atom is initially localized to a single site, and then released to the whole lattice.
The simulation is performed in the real space, despite that our analytical investigation is based
on the Bose-Hubbard model.
In our simulations, the energy, time and frequency are rescaled with
respect to the recoil energy ${{E}_{R}}\equiv \frac{{{\hbar }^{2}}{{k}^{2}}}{2M}\ $, that's to say,
the unit of the energy, time and
frequency are $E_R$, $E_R/\hbar$ and $\hbar/E_R$, respectively.
\subsubsection{The Quadratic case}
We firstly simulate the quadratic GBO, of which the external potential is composed of the linear
 and quadratic gradients. The simulation results are shown in figure \ref{fig:2}.
In figure \ref{fig:2}(a) and \ref{fig:2}(b), we compare the density oscillation of the
standard BO and the quadratic GBO. Comparing to BO, the density profile of the quadratic GBO
still maintains a periodic oscillation around the initially occupied site, while the symmetry
between the left and right wings of the profile breaks. Roughly speaking, in both wings the
propagation downwards is faster than upwards the gradient, which results in a tilted structure
of the density profile as shown in figure \ref{fig:2}(b). This parity symmetry breaking induces
 the oscillation of the center of mass, calculated by $\left\langle x \right\rangle \ $.
The center of mass remains still in the initial site during
BO, while oscillates around the initial site in the quadratic GBO,
as indicated by the dotted lines in figures \ref{fig:2}(a) and
\ref{fig:2}(b).

More importantly, the integrated spectrum $\rho \left( \omega  \right)\ $ presents more significant
difference between BO and the quadratic GBO, as compared in figures \ref{fig:2}(c) and
\ref{fig:2}(d). In both figures, we plot $\rho \left( \omega  \right)\ $ around ${{\omega }_{B}}$,
and it can be seen that in the spectrum of BO, only a single frequency peak arises located at
${{\omega }_{B}}\ $, while in the spectrum of the quadratic GBO, a series of peaks arise,
of which the spacings between neighbor peaks are the same. The neighbor spacings take exactly the
value $\delta \omega ={2{{\pi }^{2}}{{V}_{2}}}/{\hbar {{k}^{2}}}\ $ as derived in Eq. (\ref{Eq:3}).
The equidistant frequency splitting in the presence of the quadratic term is one
major result of this work.

In order to further verify the equidistant behavior in the spectrum of the quadratic GBO,
we numerically calculate the neighbor spacings over a relatively wide range of ${{V}_{2}}\ $,
and compare the results to the analytical value of $\delta \omega \ $, as shown in figure
\ref{fig:3}. In figure \ref{fig:3}, the neighbor spacings between the central four peaks in
the spectrum are plotted as a function of ${{V}_{2}}\ $, together with
the analytical value of $\delta \omega \ $.
It is clearly seen from figure \ref{fig:3} that all the spacings are lying on top of each and
match well with $\delta \omega \ $. Figure \ref{fig:3} verifies the equidistant behavior in the
spectrum of the quadratic GBO, and confirms the potential use of the quadratic GBO for the
$\textit{in-situ} $ measurement of the quadratic external potentials.
\subsubsection{Effects of higher order gradients}
In this section we numerically simulate the GBO under the external
potentials, which are composed of gradients of arbitrarily high
orders. The external potential of the general form can model complicated
forces, such as the van der Waals and the Casimir-Polder
forces. We then take the Casimir force as an example to demonstrate the GBO in the presence of
higher order gradients, and the corresponding potential is taken as the thermal Casimir-Polder
plus the gravity potential $Mgx$, which models the
interaction between atom and the dielectric surface in the presence of
the gravity field.
The linear gravity field is much stronger than the Casimir-Polder
potential, which guarantees the linear gradient dominant in the external
potential, as required in the analytical derivation.

The choice of parameters in the numerical simulations follows that in works \cite{Ant04,Ant05},
which corresponds to the
specific interaction between ${}^{40}K$ and a sapphire surface with $\epsilon_0=9.4$ at
temperature $T=300K$. The atoms are initially loaded in the $13$-th site of a lattice with
spatial period $873 nm$,
of which the local minimum is around $5.7\mu m$. The results with
lattice heights $8\ E_R$ and $12\ E_R$ are shown in figure \ref{fig:4}.
Figures \ref{fig:4}(a) and (b) present the temporal density oscillation and the
density spectrum in a lattice of height $8 E_R$. In the density oscillation, a periodic
oscillatory behavior is observed, and
more importantly, a fine structure composed of roughly four peaks
arises in the spectrum.
A qualitative agreement between the numerically calculated spectrum
and the analytical prediction by the first-order perturbation
treatment indicates that the locations of the frequency peaks are
linearly dependent on the strength of the Casimir-Polder potential,
which guarantees a potential use of the GBO spectrum for measuring
the strength of the thermal force.
Increasing the lattice barrier to $12 E_R$, the density
oscillation shrinks to fewer sites, as shown in figure \ref{fig:4}(c), and
the corresponding spectrum in figure \ref{fig:4}(d) is also reduced to
fewer peaks, which reveals the relation between the spannings in the
real space and in the spectrum. Figure \ref{fig:4}, in general,
indicates that the spectrum of the GBO with carefully chosen lattice
parameters can present fine structures, which encodes the detailed
information of the external force.

It is worth pointing out that,
the standard BO has been proposed \cite{Ant05} for measuring the
linear component of the Casimir-Polder potential, with ultracold
lattice atoms subjected to both the gravity and the Casimir force.
The proposal assumed that under the condition that the Casimir
force is much weaker than the gravity force, the effect of the
Casimir force to the spectrum of the BO is merely the shift of
the prime frequency by the linear component of the Casimir force,
and the shift can be used to measure the strength of the linear
component and consequently the Casimir force.
Our simulation, however, indicates that the Casimir force,
even though much weaker than the linear gravity force, can bring
in fine structures to the spectrum. Under a limited
spectrum resolution in experiments, the fine structure could be
seen a broadening of the prime frequency, of which the width is
around $10^{-4}\omega_B$. Then in this case, the fine structure
leads to an intrinsic error to the proposal.
On the other hand, in the case that the fine structure can be
resolved in experiments with improved resolution of the measured
spectrum, it can be used to decode more information of the external force.
For instance, the ratio between the spacings of the peaks in the fine
structure can work as a fingerprint of the external potentials, and
be used to distinguish one type of potential to another. The exact
value of these peaks could further determine the strength of the potential.

\section{DISCUSSION} \label{section:III}
In this work, we have theoretically investigated the generalized Bloch oscillations (GBOs)
of non-interacting ultracold atoms in optical lattices in the presence of external potentials
with higher order gradients.
The results demonstrate that the spectrum of GBO presents fine structures
of multiple series of frequency peaks, which encodes the properties, such as the strength of
different gradients, of the external potential. With
the fast progress of the experimental techniques, the fine structure could
be within the reach of experiments in the recent future.

The fine structure gives rise to a two-fold consequence to the
application of using the BO (or GBO) for measuring the external force.
For one thing, in the case of a limited resolution of the experimentally
measured spectrum, the fine structures could manifest
as a broadening of the Bloch frequencies,
and brings in errors to the measured results.
For another, if the fine structure could be resolved in experiments, it
could supply more information than that given by BO. The spectrum of BO or
approximated BO can only tell the strength of the linear gradient of
the external potentials, while the fine structure could be used as a fingerprint to identify
the type of external potential.
The GBO under higher order gradients then possesses the potential use of
the lattice atoms tunneling for measuring not only the linear, but the higher order gradients,
which offers new possibilities to measure complicated forces with ultracold atom
ensembles \cite{Wol07,Har05,Suk93,Lan96,Shi01,Lin04,Lem05,Cas48,Ant05,Ant04,Har03,McG04,Dim03}.
\section*{ACKNOWLEDGMENTS}
The authors would like to acknowledge Y. Chang and T. Shi for inspiring
discussions. This work was supported by the National Natural Science
Foundation of China (Grants Nos. 11625417, No. 11604107, No. 91636219 and No. 11727809).
\begin{appendix}
\section{THE ANALYTIC DERIVATION OF EQUIDISTANT SPLITTING IN SPECTRUM MAP}
In the appendix, we derive the dependence of the characteristic peaks in the spectrum of GBO, by applying perturbation treatments to the Wannier-Stark states. Since the GBO only spans over a few sites around the initial one, in which the atoms are initially loaded to, we can adapt the tight-binding Bose-Hubbard Hamiltonian to describe the system, which reads:
\begin{equation}
\hat{H}=t\sum\limits_{i}{\left( \hat{b}_{i}^{\dagger }{{{\hat{b}}}_{i+1}}+h.c. \right)}+\sum\limits_{i}{{{U}_{i}}\hat{b}_{i}^{\dagger }{{{\hat{b}}}_{i}}}\ ,\label{Eq:A1}
\end{equation}
where $\hat{b}_{i}^{\left( \dagger  \right)}\ $ is the annihilation (creation) operator on the ${i} $-th lattice site, with ${t} $  and ${{U}_{i}}$ denoting the strength of nearest-neighbor hopping and the on-site potential, respectively. In the following, we approximate
${{U}_{i}}$ taking the potential of the original local minimum
of the $i$-th site in the optical lattice, that's
${{U}_{i}}=
\sum_{\alpha\geq 1}\tilde{V}_\alpha i^\alpha \left( \pi/k\right)^\alpha
\equiv\sum_{\alpha\geq 1}U_\alpha i^\alpha$. In the equations of the
appendix, the summation runs over all integers, except explicitly
clarified.

To apply the perturbation treatment, we consider the case that the gradients with order higher than one are small and correspond to the perturbative terms to the linear gradient. Then the unperturbed eigenstates are the Wannier-Stark states in a linearly tilted lattice, which reads:
\begin{equation}
{{\left| n \right\rangle }^{\left( 0 \right)}}=\sum\limits_{\Delta }{{{J}_{\Delta }}\left( \frac{2t}{{{{U}}_{1}}} \right)\left| n+\Delta  \right\rangle }\ ,\label{Eq:A2}
\end{equation}
where ${{J}_{\Delta }}\left( x \right)\ $ is the $\Delta \ $-th order Bessel function of the first kind, and $\left| n \right\rangle \ $ refers to the lowest Wannier state in the ${n} $-th site. The corresponding unperturbed eigenenergy is:
\begin{equation}
E_{n}^{\left( 0 \right)}={{E}_{0}}+n{{U}_{1}}\ .\label{Eq:A3}
\end{equation}

The first order corrections to the eigenstates read:
\begin{equation}
{{\left| n \right\rangle }^{\left( 1 \right)}}=\sum\limits_{\Delta }{\left[ \left( \sum_{\alpha \geq 1}{{{c}_{\alpha }}} \right){{\left| n+\Delta  \right\rangle }^{\left( 0 \right)}} \right]} , \label{Eq:A4}
\end{equation}

\begin{equation}
{{c}_{\alpha}}=-\frac{{{{U}}_{\alpha }}}{\Delta {{{U}}_{1}}}\sum\limits_{\delta }{\left( {{J}_{\delta }}{{J}_{\delta -\Delta }}{{\left( n+\delta  \right)}^{\alpha }} \right)}\ , \label{Eq:A5}
\end{equation}
where ${{c}_{\alpha }}\ $ is the contribution of the $\alpha \ $-th gradients with $\alpha \ge 2\ $. The first order corrections to the eigenenergies are:
\begin{equation}
E_{n}^{\left( 1 \right)}=\sum\limits_{\alpha \geq 1 }{{{{U}}_{\alpha }}\left( \sum\limits_{\Delta }{{{\left( n+\Delta  \right)}^{\alpha }}J_{\Delta }^{2}} \right)}\ ,\label{Eq:A6}
\end{equation}

Provided the eigenstates and corresponding eigenenergies up to the first order correction, $\left| n \right\rangle ={{\left| n \right\rangle }^{\left( 0 \right)}}+{{\left| n \right\rangle }^{\left( 1 \right)}}\equiv \sum\limits_{\Delta }{{{C}_{\Delta }}\left( n \right)\left| n+\Delta  \right\rangle }\ $ and ${{E}_{n}}=E_{n}^{\left( 0 \right)}+E_{n}^{\left( 1 \right)}\ $, we can derive the spectrum of the one-body density, with a focus on the position of the peaks in the spectrum. Consider that the atom is initially loaded in the ${j} $-th site, and we focus on the density spectrum of the oscillation in the same site, of which the characteristic frequencies reads:
\begin{equation}
\omega \left( n,m \right)=\frac{1}{\hbar }\sum\limits_{\alpha \ge 1}{{{U}_{\alpha }}}\sum\limits_{l=0}^{\alpha }{\left[ C_{\alpha }^{l}\left( {{n}^{\alpha -l}}-{{m}^{\alpha -l}} \right)\left\langle {{\delta }^{l}} \right\rangle  \right]}\ ,
\label{Eq:A8}
\end{equation}
in which $\omega\left( n,m \right)$ denotes the characteristic frequency
related to the energy difference between the $n$- and $m$-th eigenstates,
and $\left\langle {{\delta }^{l}} \right\rangle =\sum\limits_{\delta }{J_{\delta }^{2}{{\delta }^{l}}}\ $ denotes the $l\ $-th order variance.
One can further show that $\langle {{\delta }^{l}} \rangle=0$ for the odd
$l$.

Particularly, the characteristic frequencies around $\omega_B$, which we
focus in the maintext, are given by $\omega(n,n-1)$.
\end{appendix}


\begin{thebibliography}{99}
\bibitem{Blo08} I. Bloch, J. Dalibard and W. Zwerger, Rev. Mod. Phys. {\bf 80}, 885 (2008).

\bibitem{Jak05} D. Jaksch and P. Zoller, Annals of Physics, {\bf 315}, 52 (2005).

\bibitem{Zen34} C. Zener and R. H. Fowler, Proceedings of the Royal Society of London. Series A, Containing Papers of a Mathematical and Physical Character {\bf 145}, 523 (1934).

\bibitem{Dah96} M. B. Dahan, E. Peik, J. Reichel, Y. Castin, and C. Salomon, Phys. Rev. Lett. {\bf 76}, 4508 (1996).

\bibitem{Mor01} O. Morsch, J. H. M\"{u}ller, M. Cristiani, D. Ciampini, and E. Arimondo, Phys. Rev. Lett. {\bf 87}, 140402 (2001).

\bibitem{Pri98} G. A. Prinz, Science {\bf 282}, 1660 (1998).

\bibitem{Bat04} R. Battesti, P. Clad\'{e}, Sa\"{i}da Guellati-Kh\'{e}lifa, C. Schwob, B. Gr\'{e}maud, F. Nez, L. Julien, and F. Biraben, Phys. Rev. Lett. {\bf 92}, 253001 (2004).

\bibitem{Cla06} P. Clad\'{e}, Estefania de Mirandes, M. Cadoret, Sa\"{i}da Guellati-Kh\'{e}lifa, C. Schwob, F. Nez, L. Julien, and F. Biraben, Phys. Rev. Lett. {\bf 96}, 033001 (2006).

\bibitem{Fer06} G. Ferrari, N. Poli, F. Sorrentino, and G. M. Tino, Phys. Rev. Lett. {\bf 97}, 060402 (2006).

\bibitem{Gus08} M. Gustavsson, E. Haller, M. J. Mark, J. G. Danzl, G. Rojas-Kopeinig, and H.-C. N\"{a}gerl, Phys. Rev. Lett. {\bf 100}, 080404 (2008).

\bibitem{Gei18} Z. A. Geiger, K. M. Fujiwara, K. Singh, R. Senaratne, S. V. Rajagopal, M. Lipatov, T. Shimasaki, R. Driben, V. V. Konotop, T. Meier, and D. M. Weld, Phys. Rev. Lett. {\bf 120}, 213201 (2018).

\bibitem{Fat08} M. Fattori, C. D¡¯Errico, G. Roati, M. Zaccanti, M. Jona-Lasinio, M. Modugno, M. Inguscio, and G. Modugno, Phys. Rev. Lett. {\bf 100}, 080405 (2008).

\bibitem{Roa04} G. Roati, E. de Mirandes, F. Ferlaino, H. Ott, G. Modugno, and M. Inguscio, Phys. Rev. Lett. {\bf 92}, 230402 (2004).

\bibitem{Pre15} P. M. Preiss, R. Ma, M. Eric Tai, A. Lukin, M. Rispoli, P. Zupancic, Y. Lahini, R. Islam, M. Greiner, Science {\bf 347}, 1229 (2015).

\bibitem{Cla05} P. Clad\'{e}, S. Guellati-Kh\'{e}lifa, C. Schwob, F. Nez, L. Julien and F. Biraben, Europhys. Lett. {\bf 71}, 730 (2005).

\bibitem{Bre07} B. M. Breid, D. Witthaut and H. J. Korsch, New J. Phys. {\bf 9}, 62 (2007).

\bibitem{Car05} I. Carusotto, L. Pitaevskii, S. Stringari, G. Modugno, and M. Inguscio, Phys. Rev. Lett. {\bf 95}, 093202 (2005).

\bibitem{Sor09} F. Sorrentino, A. Alberti, G. Ferrari, V. V. Ivanov, N. Poli, M. Schioppo, and G. M. Tino, Phys. Rev. A {\bf 79}, (2009).

\bibitem{Wan04} J. Wan, C. Martijn de Sterke, and M. M. Dignam, Phys. Rev. B {\bf 70}, 125311 (2004).

\bibitem{Tar12} M. G. Tarallo, A. Alberti, N. Poli, M. L. Chiofalo, F.-Y. Wang, and G. M. Tino, Phys. Rev. A {\bf 86}, 033615 (2012).

\bibitem{Iva08} V. V. Ivanov, A. Alberti, M. Schioppo, G. Ferrari, M. Artoni, M. L. Chiofalo, and G. M. Tino, Phys Rev Lett. {\bf 100}, 043602 (2008).

\bibitem{Hal10} E. Haller, R. Hart, M. J. Mark, J. G. Danzl, L. Reichs\"{o}llner, and Hanns-Christoph N\"{a}gerl, Phys Rev Lett. {\bf 104}, 200403 (2010).

\bibitem{Pol11} N. Poli, F.-Y. Wang, M. G. Tarallo, A. Alberti, M. Prevedelli, and G. M. Tino, Phys Rev Lett. {\bf 106}, 038501 (2011).

\bibitem{Dia13} E. D\'{i}az, A. G. Mena, K. Asakura, and C. Gaul, Phys. Rev. A {\bf 87}, 015601 (2013).

\bibitem{Wal10} S. Walter, D. Schneble, and A. C. Durst, Phys. Rev. A {\bf 81}, 033623 (2010).

\bibitem{Dug16} L. Duggen, L. C. Lew Yan Voon, B. Lassen and M. Willatzen, J Phys Condens Matter {\bf 28}, 155301 (2016).

\bibitem{Sch08} T. Schulte, S. Drenkelforth, G. Kleine B\"{u}ning, W. Ertmer, J. Arlt, M. Lewenstein, and L. Santos, Phys. Rev. A {\bf 77}, 023610 (2008).

\bibitem{Sto15} J. Stockhofe and P. Schmelcher, Phys. Rev. A {\bf 91}, 023606 (2015).

\bibitem{Glu02} M. Gl\"{u}ck, A. R. Kolovsky, H. J. Korsch, Phys. Rep. {\bf 366}, 103 (2002).

\bibitem{Wan60} G. H. Wannier, Phys. Rev. {\bf 117}, 432 (1960).

\bibitem{Wol07} P. Wolf, P. Lemonde, A. Lambrecht, S. Bize, A. Landragin, and A. Clairon, Phys. Rev. A {\bf 75}, 063608 (2007).

\bibitem{Har05} D. M. Harber, J. M. Obrecht, J. M. McGuirk, and E. A. Cornell, Phys. Rev. A {\bf 72}, 033610 (2005).

\bibitem{Suk93} C. I. Sukenik, M. G. Boshier, D. Cho, V. Sandoghar, and E. A. Hinds, Phys. Rev. Lett. {\bf 70}, 560 (1993).

\bibitem{Lan96} A. Landragin, J. Y. Courtois, G. Labeyrie, N. Vansteenkiste, C. I. Westbrook, and A. Aspect, Phys. Rev. Lett. {\bf 77}, 1464 (1996).

\bibitem{Shi01} F. Shimizu, Phys. Rev. Lett. {\bf 86}, 987 (2001).

\bibitem{Lin04} Y. J. Lin, I. Teper, C. Chin, and V. Vuletic, Phys. Rev. Lett. {\bf 92}, 050404 (2004).

\bibitem{Lem05} P. Lemonde and P. Wolf, Phys. Rev. A {\bf 72}, 033409 (2005).

\bibitem{Cas48} H. G. B. Casimir and D. Polder, Phys. Rev. {\bf 73}, 360 (1948).

\bibitem{Ant05} M. Antezza, L. P. Pitaevskii, and S. Stringari, Phys. Rev. Lett. {\bf 95}, 113202 (2005).

\bibitem{Ant04} M. Antezza, L. P. Pitaevskii, and S. Stringari, Phys. Rev. A {\bf 70}, 053619 (2004).

\bibitem{Har03} D. M. Harber, J. M. McGuirk, J. M. Obrecht, and E. A. Cornell, J. Low Temp. Phys. {\bf 133}, 229 (2003).

\bibitem{McG04} J. M. McGuirk, D. M. Harber, J. M. Obrecht, and E. A. Cornell, Phys. Rev. A {\bf 69}, 062905 (2004).

\bibitem{Dim03} S. Dimopoulos and A. A. Geraci, Phys. Rev. D {\bf 68}, 124021 (2003).

\end{thebibliography}
\end{document}